\documentclass[12pt]{iopart}
\usepackage{iopams}  
\begin{document}

\title[BBGKY hierarchy - new truncation scheme]{A new truncation scheme for BBGKY hierarchy: conservation of energy and time reversibility}

\author{S Siva Nasarayya Chari$^1$, Ramarao Inguva$^2$ and K P N Murthy$^3$}
\address{$^1$School of Physics, University of Hyderabad, Hyderabad 500046, India.}
\ead{ssnchari@gmail.com} 

\vspace{3pt}
\address{$^2$Visiting Professor, Centre for Modeling, Simulation and Design (CMSD),\\
	University of Hyderabad, Hyderabad 500046, India.}
\ead{ringuvausa@gmail.com} 

\vspace{3pt}
\address{$^3$Adjunct Professor, Manipal Centre for Natural Sciences (MCNS),\\ Manipal University, Manipal 576104, India.}
\ead{k.p.n.murthy@gmail.com}

\vspace{10pt}

\begin{abstract}
We propose a new truncation scheme for Bogoliubov-Born-Green-Kirkwood-Yvon (BBGKY) hierarchy. We approximate the three particle distribution function $f_{3}(1,2,3,t)$ in terms of $f_{2}(1,2,t)$, $f_{1}(3,t)$ and two point correlation functions $\left\lbrace g_{2}(1,3,t), g_{2}(2,3,t)\right\rbrace $. Further $f_{2}$ is expressed in terms of $f_{1}(1,t)$ and $g_{2}(1,2,t)$ to close the hierarchy, resulting a set of coupled kinetic equations for $f_{1}$ and $g_{2}$. In this paper we show that, for velocity independent correlations, the kinetic equation for $f_{1}$ reduces to the model proposed by Martys[Martys N S 1999 \textit{IJMPC} \textbf{10} 1367-1382]. In the steady state limit, the kinetic equation for $g_{2}$ reduces to Born-Green-Yvon (BGY) hierarchy for homogeneous density. We also prove that the present scheme respects the energy conservation and under specific circumstances, time symmetry \textit{i.e.,} $\displaystyle \frac{dH(t)}{dt} = 0$ where $H(t)$ refers to the Boltzmann's H-function.
\end{abstract}

%
\vspace{2pc}
%
\maketitle
%
%

\section{Introduction}\label{sec:intro}

Consider a system of ~$N$~ interacting gas molecules present in a volume ~$V$~ under the influence of an external potential ~$U(\vec{r})$. From Liouville's theorem one can say that, the ~$s$-particle phase space distribution function ~${f_{s}(1,2,\cdots,s,t)},~{\left(1 \leq s \leq N-1 \right)}$~ of such a system is related to the ~$(s+1)$-particle distribution function through BBGKY hierarchy~\cite{BBGKY_ref1,BBGKY_ref2} of equations,
\begin{eqnarray}
	\frac{\partial f_s}{\partial t} + \sum_{i=1}^{s}\frac{\vec{p}_{i}}{m} \cdot \frac{\partial f_s}{\partial \vec{r}_{i}} + \sum_{i=1}^{s}\vec{F}_{i} \cdot \frac{\partial f_s}{\partial \vec{p}_{i}} &-& \sum_{i=1}^{s} \frac{\partial f_s}{\partial \vec{p}_{i}} \cdot \frac{\partial }{\partial \vec{r}_{i}} \left( \frac{1}{2} {\sum_{j=1 (\neq i)}^{s}} \phi_{ij} \right) = \nonumber \\
	& =& \sum_{i=1}^{s} \int \frac{\partial f_{s+1}}{\partial \vec{p}_{i}} \cdot \frac{\partial \phi_{i,s+1}}{\partial \vec{r}_{i}} ~d\omega_{s+1}~.
	\label{bbgky_hierarchy}
\end{eqnarray}
In the above, $\phi_{ij}\left[=\phi(|\vec{r_{i}}-\vec{r_{j}}|)\right]$ is the pair interaction potential between $i^{th}$ and $j^{th}$ particles, ~$\vec{F_{i}} = -\vec{\nabla}U(\vec{r}_{i})$~ and
\begin{equation}
f_{s}(1, 2,\cdots,s,t) = \frac{N!}{(N-s)!}\int{\prod_{r=s+1}^{N}{d\omega_{r}}~\rho(1,2,\cdots,N,t)}~
\end{equation}
where $\rho(1,2,\cdots,N,t)$ is phase space density of the total system. One need to approximate the higher order distribution functions $f_{s+1},~{(1~\leq~s~\leq~N-1)}$ to obtain a closed set of equations for $f_{s}$. Solution of these kinetic equations holds importance in understanding the transport processes in dense gaseous systems. There were recent investigations in this direction {\it e.g.},~\cite{XuSucBogh_MCS2006,Mar_IJMPC1999}, where hierarchy is truncated at $f_{2}$ to derive a model kinetic equation for $f_{1}$. In this context, we propose a new closure scheme for the hierarchy at the level of three 
particle distribution function $f_3(1,2,3,t)$. The scheme and respective calculation is discussed in the following section.

\section{New scheme of closure for hierarchy}\label{sec:2}

From \eref{bbgky_hierarchy}, we get for $s=1$ and $s=2$ as
\begin{equation} 
	\frac{\partial f_{1}(1)}{\partial t} + \vec{v}_{1} \cdot \vec{\nabla}_{r_1}f_{1}(1) = \int \frac{\partial \phi_{12}}{\partial \vec{r}_{1}} \cdot \frac{\partial f_{2}(1,2)}{\partial \vec{v}_{1} }~d^{3}r_{2}~d^{3}v_{2}
	\label{bbgky_s_1_f0}
\end{equation}
and
\begin{eqnarray}
	\frac{\partial f_{2}(1,2)}{\partial t} &+& \vec{v}_{1} \cdot \vec{\nabla}_{r_1}f_{2} + \vec{v}_{2} \cdot \vec{\nabla}_{r_2}f_{2} +   \frac{\vec{F}_{12}}{m} \cdot (\vec{\nabla}_{v_1} - \vec{\nabla}_{v_2})f_{2} = \nonumber \\ & =& \int \left[ \frac{\partial \phi_{13}}{\partial \vec{r}_{1}} \cdot \frac{\partial f_{3}(1,2,3)}{\partial \vec{v}_{1}} + \frac{\partial \phi_{23}}{\partial \vec{r}_{2}} \cdot \frac{\partial f_{3}(1,2,3)}{\partial \vec{v}_{2}} \right] ~d^{3}r_{3}~d^{3}v_{3} 
	\label{bbgky_s_2_f0}
\end{eqnarray}
respectively, in absence of external forces $(\vec{F}={\bf 0})$. The above set of equations is not closed as $f_{3}$ is not known. We approximate $f_{3}$ in terms of its lower order distribution functions,
\begin{eqnarray}
&~&f_{3}(1,2,3) = f_{2}(1,2)~f_{1}(3)~g_{2}(1,3)~g_{2}(2,3)~ 
\label{scheme_123} \\
\mbox{and}\hspace*{2cm} &~&f_{2}(1,2) = f_{1}(1)~f_{1}(2)~g_{2}(1,2)~.
\label{scheme_12}
\end{eqnarray} 
From \eref{scheme_12} and \eref{bbgky_s_1_f0}, we get
\begin{eqnarray} 
	\frac{\partial f_{1}(1)}{\partial t} + \vec{v}_{1} \cdot \vec{\nabla}_{r_1}f_{1}(1) &=& \int \frac{\partial \phi_{12}}{\partial \vec{r}_{1}} \cdot \frac{\partial }{\partial \vec{v}_{1} } [f_{1}(1)~f_{1}(2)~g_{2}(1,2)]~d^{3}r_{2}~d^{3}v_{2}~, \nonumber \\[2mm]
	&=& \frac{\partial f_{1}(1)}{\partial \vec{v}_{1}} \cdot \int \frac{\partial \phi_{12}}{\partial \vec{r}_{1}}~f_{1}(2)~g_{2}(1,2)~d^{3}r_{2}~d^{3}v_{2} ~+ \nonumber \\[2mm]
	&+& f_{1}(1)~\int \frac{\partial \phi_{12}}{\partial \vec{r}_{1}} \cdot \frac{\partial g_{2}(1,2)}{\partial \vec{v}_{1}}~f_{1}(2)~d^{3}r_{2}~d^{3}v_{2}~.
\end{eqnarray}
The above equation can be rewritten as,
\begin{eqnarray} 
	\left[ \frac{\partial }{\partial t} + \vec{v}_{1} \cdot \vec{\nabla}_{r_1} + \vec{a}_{int}^{(1)}(1) \cdot \frac{\partial }{\partial \vec{v}_{1}} \right] \ln(f_{1}(1)) = 
(\vec{\nabla}_{r^{+}_{1}} \cdot \vec{\nabla}_{v^{+}_{1}}) \Phi_{eff}^{(1)}(1^{+},2)~, \label{eq:A}
\end{eqnarray}
where
\begin{eqnarray} 
	&~&\vec{a}_{int}^{(1)}(1) = -\int \frac{\partial \phi_{12}}{\partial \vec{r}_{1}}~f_{1}(2)~g_{2}(1,2)~d^{3}r_{2}~d^{3}v_{2} \\[2mm]
	\mbox{and}\hspace*{1.5em} &~&\Phi_{eff}^{(1)}(1^{+},2) = \int \phi_{1^{+}2}~g_{2}(1^{+},2)~f_{1}(2)~d^{3}r_{2}~d^{3}v_{2}~. 
\end{eqnarray}
From \eref{bbgky_s_2_f0} and \eref{scheme_123}, we write, 
\begin{eqnarray}
	\frac{\partial f_{2}(1,2)}{\partial t} + \vec{v}_{1} \cdot \vec{\nabla}_{r_1}f_{2}(1,2) &+& \vec{v}_{2} \cdot \vec{\nabla}_{r_2}f_{2}(1,2) +   \frac{\vec{F}_{12}}{m} \cdot (\vec{\nabla}_{v_1} - \vec{\nabla}_{v_2})f_{2}(1,2) = \nonumber \\[2mm]
	&=&\int \left[ C_{1} + C_{2} \right] ~d^{3}(3)  \label{bbgky_s_2_f0_rhs} 
\end{eqnarray}
where
\begin{eqnarray} 
C_{1} &=& \frac{\partial \phi_{13}}{\partial \vec{r}_{1}} \cdot \frac{\partial}{\partial \vec{v}_{1}}[f_{2}(1,2)~f_{1}(3)~g_{2}(1,3)~g_{2}(2,3)]~, \nonumber \\[2mm]
C_{2} &=& \frac{\partial \phi_{23}}{\partial \vec{r}_{2}} \cdot \frac{\partial}{\partial \vec{v}_{2}}[f_{2}(1,2)~f_{1}(3)~g_{2}(1,3)~g_{2}(2,3)]~, 
\nonumber
\end{eqnarray} 
and $d^{3}(3) = d^{3}r_{3}~d^{3}v_{3}$ respectively.
The above equation can be further simplified as, 
\begin{eqnarray}
	\left[ S_{12}^{\prime} \right] \ln(g_{2}(1,2)) = &-&a_{int}^{eff}(1^{+}2) \cdot \vec{\nabla}_{v_{1}} \ln(f_{1}(1)) - \nonumber \\[2mm] 
	&-&a_{int}^{eff}(12^{+}) \cdot \vec{\nabla}_{v_{2}} \ln(f_{1}(2)) + \nonumber \\[2mm]
	&+& (\vec{\nabla}_{r^{+}_{1}} \cdot \vec{\nabla}_{v^{+}_{1}})\left[ \Psi^{(2)}(1^{+}2) - \Phi_{eff}^{(1)}(1^{+}2) \right] + \nonumber \\[2mm]
	&+& (\vec{\nabla}_{r^{+}_{2}} \cdot \vec{\nabla}_{v^{+}_{2}})\left[ \Psi^{(2)}(12^{+}) - \Phi_{eff}^{(1)}(12^{+}) \right]~.
	\label{eq:B}
\end{eqnarray}
Here, 
\begin{eqnarray}
	\left[ S_{12}^{\prime} \right] = \frac{\partial}{\partial t} &+& \vec{v}_{1} \cdot \vec{\nabla}_{r_{1}} + \vec{v}_{2} \cdot \vec{\nabla}_{r_{2}} + a_{int}^{(2)}(1^{+}2) \cdot \vec{\nabla}_{v_{1}} + a_{int}^{(2)}(12^{+}) \cdot \vec{\nabla}_{v_{2}}~,
	\label{eq:1.30} \\[2mm]
	\Psi^{(2)}(1^{+}2) &=& \int \phi(1^{+}3)~g_{2}(1^{+}3)~f_{1}(3)~g_{2}(2,3)~d^{3}(3)~, \\[2mm]
	\Psi^{(2)}(12^{+}) &=& \int \phi(2^{+}3)~g_{2}(2^{+}3)~f_{1}(3)~g_{2}(1,3)~d^{3}(3)~, \\[2mm]
	a_{int}^{eff}(1^{+}2) &=& a_{int}^{(2)}(1^{+}2) - a_{int}^{(1)}(1)~, \\[2mm]
	a_{int}^{eff}(12^{+}) &=& a_{int}^{(2)}(12^{+}) - a_{int}^{(1)}(2)~.\\[2mm]
\end{eqnarray}
where 
\begin{eqnarray}
	a_{int}^{(2)}(1^{+}2) &=& -\int \frac{\partial \phi_{13}}{\partial \vec{r}_{1}} ~f_{1}(3) ~g_{2}(1,3) ~g_{2}(2,3) ~d^{3}(3)~, \\[2mm]
	a_{int}^{(2)}(12^{+}) &=& -\int \frac{\partial \phi_{23}}{\partial \vec{r}_{2}} ~f_{1}(3) ~g_{2}(1,3) ~g_{2}(2,3) ~d^{3}(3)~.
\end{eqnarray} 
Equations~(\ref{eq:A})~ and ~(\ref{eq:B})~ together form a closed set of equations. A numerical solution of these coupled integro-differential equations can be obtained for a given initial condition on $f_{1}$ and $g_{2}$. Unlike the earlier models~\cite{Mar_IJMPC1999,Mar_PhysA2006,Siva_JSP2014}, velocity and time dependence of $g_{2}$ is retained in the present model. Hence it yields more reliable results of time evolution of $f_{1}$. In the following subsection, we present the total energy conservation succeeded by a discussion on time behavior of Boltzmann's H-function for the present truncation scheme. 

\subsection{Conservation of total energy}

The time rate of change of local kinetic energy $(K_{loc}(\vec{r}))$ can be calculated as,
\begin{equation}
\frac{d}{dt}\left(K_{loc}(\vec{r}_{1},t)\right) = 
\int d^{3}v_{1} \left( \frac{1}{2}m\vec{v}_{1}^{2} \right)  d^{3}(2) \left\lbrace \frac{1}{m} \int \left[ \frac{\partial \phi_{13}}{\partial \vec{r}_{1}} \cdot \frac{\partial f_{3}}{\partial \vec{v}_{1}} + \frac{\partial \phi_{23}}{\partial \vec{r}_{2}} \cdot \frac{\partial f_{3}}{\partial \vec{v}_{2}} \right] d^{3}(3) \right\rbrace ~.
\label{K_loc}
\end{equation}
Upon integrating by parts with respect to $\vec{v}_{1}$ and $\vec{v}_{2}$, assuming $f_{3}$ vanishes at infinite velocities, the above equation results in 
\begin{equation}
\frac{d}{dt}\left(K_{loc}(\vec{r}_{1},t)\right) = 
-\int d^{3}\vec{v}_{1} ~d^{3}(2) ~d^{3}(3) \left[ \vec{v}_{1}\cdot\vec{\nabla}_{r_{1}}\phi_{13} \right] f_{3}(1,2,3,t)~.
\end{equation}
From the above, we can write,
\begin{equation}
\frac{d}{dt}\left(K_{tot}(t)\right) = 
-\int d^{3}(1)~d^{3}(3) \left[ \vec{v}_{1}\cdot\vec{\nabla}_{r_{1}}\phi_{13} \right] f_{2}(1,3,t)~,
\label{K_tot_time}
\end{equation}
where 
\begin{equation}
K_{tot}(t) = \int K_{loc}(\vec{r},t) ~d^{3}r~.
\end{equation}
Assuming only two-body interactions, the time rate of change of total potential energy can be written as,
\begin{equation}
\frac{d}{dt}\left(\Phi_{tot}(t)\right) = 
\frac{1}{2} \int \phi_{12} ~\frac{\partial f_{2}}{\partial t}~d^{3}(1)~d^{3}(2)~,
\label{PE_dt}
\end{equation}
where 
\begin{equation}
\Phi_{tot}(t) = \frac{1}{2} \int \phi_{12} ~f_{2}(1,2,t)~d^{3}(1)~d^{3}(2)~
\end{equation}
is the total potential energy of the system at any instant of time $t$. From \eref{bbgky_s_2_f0} and \eref{PE_dt}, we write
\begin{equation}
\frac{d}{dt}\left(\Phi_{tot}(t)\right) =
\frac{1}{2} \int \phi_{12} \left\lbrace -\left[ \Theta \right]f_{2} 
+ \int \left( \frac{\partial \phi_{13}}{\partial \vec{r}_{1}} \cdot \frac{\partial f_{3}}{\partial \vec{v}_{1}} + \frac{\partial \phi_{23}}{\partial \vec{r}_{2}} \cdot \frac{\partial f_{3}}{\partial \vec{v}_{2}} \right) d^{3}(3) \right\rbrace d^{3}(1)~d^{3}(2)~.
\end{equation}
where 
\begin{equation}
\Theta = \vec{v}_{1}\cdot\vec{\nabla}_{r_{1}} + \vec{v}_{2}\cdot\vec{\nabla}_{r_{2}} + \frac{\vec{F}_{12}}{m}\cdot(\vec{\nabla}_{v_{1}} - \vec{\nabla}_{v_{2}})~.
\end{equation}
Upon integrating by parts with respect to $\vec{v}_{1}$, $\vec{v}_{2}$, $\vec{r}_{1}$ and $\vec{r}_{2}$, assuming the distribution function vanishes at $v = \pm~\infty$ and $r = \pm~\infty$~, we obtain
\begin{equation}
\frac{d}{dt}\left(\Phi_{tot}(t)\right) = \int d^{3}(1)~d^{3}(2)~\left( \vec{v}_{1}\cdot\vec{\nabla}_{r_{1}} \phi_{12} \right)f_{2}(1,2,t)~.
\label{PE_tot_time}
\end{equation}
Hence, from \eref{K_tot_time} and \eref{PE_tot_time}, we notice that
\begin{equation}
\frac{d}{dt}\left( K_{tot}(t) \right) = - \frac{d}{dt}\left( \Phi_{tot}(t) \right)~,
\end{equation}
and hence the total energy is conserved. 

\subsection{H-function and time reversibility}

We have the Boltzmann's H-function defined as,
\begin{equation}
H(t) = \int d^{3}(1)~ f_{1}(1,t) ~\ln f_{1}(1,t)~.
\end{equation}
From \eref{bbgky_s_1_f0}, the time rate of change of $H(t)$ can be written as, 
\begin{eqnarray}
\frac{dH(t)}{dt} &=& -\int d^{3}(1)~d^{3}(2)\left[ \vec{\nabla}_{r_{1}}\phi_{12}\cdot\vec{\nabla}_{v_{1}}\ln f_{1}(1,t) \right] f_{2}(1,2,t)~
\label{H_st_1} \\
&=& -\int d^{3}(1)~d^{3}(2) \left[\vec{\nabla}_{r_1}\phi_{12}\cdot\vec{\nabla}_{v_1}f_{1}(1,t)\right]f_{1}(2,t)~g_{2}(1,2,t)~. 
\label{H_st_2} \\
&~& ~~~~\mbox{(using \eref{scheme_12})}
\end{eqnarray}
Now, assuming the density is homogeneous, we can write
\begin{equation}
g_{2}(1,2,t) = g_{2}(r_{12},v_{12},t)~,
\label{g2_hom}
\end{equation}
where $r_{12}=|\vec{r}_{2}-\vec{r}_{1}|$ and $v_{12}=|\vec{v}_{2}-\vec{v}_{1}|$ respectively. Hence from \eref{H_st_2} and \eref{g2_hom}, 
\begin{equation}
\frac{dH(t)}{dt} =
\int d^{3}r_{1}~\hat{r}_{1}\cdot\int d^{3}v_{1}~\hat{v}_{1}~\frac{\partial f_{1}(1,t)}{\partial v_{1}}~\left[ I_{12} \right] 
\label{dH_dt_0}
\end{equation}
where
\begin{equation}  
\left[ I_{12} \right] =  \int~d^{3}r_{12}~d^{3}v_{12}~f_{1}(2,t)
~g_{2}(r_{12},v_{12},t)~\frac{\partial \phi_{12}}{\partial r_{12}}~. \\
\end{equation}
Since the first integral in \eref{dH_dt_0} is the vector sum of all possible directions, we get
\begin{equation}
\frac{dH(t)}{dt} = 0~.
\end{equation} 
Since we have a closed equation for $f_{2}(1,2)$, let us define the $H$-function as,
\begin{equation}
H_{2}(t) = \int d^{3}(1)~d^{3}(2)~f_{2}(1,2,t) \ln f_{2}(1,2,t)~.
\end{equation}
From \eref{bbgky_s_2_f0}, we can write 
\begin{equation}
\frac{dH_{2}(t)}{dt} = \int d^{3}(1)~d^{3}(2) \int d^{3}(3) \left[ \vec{\nabla}_{r_1}\phi_{13}\cdot\vec{\nabla}_{v_1}f_{3}+\vec{\nabla}_{r_2}\phi_{23}\cdot\vec{\nabla}_{v_2}f_{3} \right] \ln f_{2}(1,2)~.
\end{equation} 
Upon integrating by parts with respect to $\vec{v}_{1}$ and $\vec{v}_{2}$, the above equation becomes
\begin{equation}
\frac{dH_{2}}{dt} = -\int d^{3}(1)~d^{3}(2)~d^{3}(3) \left[ \vec{\nabla}_{r_1}\phi_{13}\cdot\vec{\nabla}_{v_1}\ln f_{2}(1,2)+\vec{\nabla}_{r_2}\phi_{23}\cdot\vec{\nabla}_{v_2}\ln f_{2}(1,2) \right] f_{3}~.
\end{equation}
From \eref{scheme_123} and \eref{scheme_12}, we get
\begin{equation}
\frac{dH_{2}}{dt} = -\int d^{3}(1)~d^{3}(2)~d^{3}(3) \left[ \vec{\nabla}_{r_1}\phi_{13}\cdot\vec{\nabla}_{v_1}\ln f_{1}(1)+\vec{\nabla}_{r_2}\phi_{23}\cdot\vec{\nabla}_{v_2}\ln f_{1}(2) \right] f_{3}~.
\end{equation}
By integrating out the $3^{rd}$ particle that is not necessary, the above equation reduces to
\begin{equation}
\frac{dH_{2}}{dt} = -2 \int d^{3}(1)~d^{3}(3) \left[ \vec{\nabla}_{r_1}\phi_{13}\cdot\vec{\nabla}_{v_1}\ln f_{1}(1) \right] f_{2}(1,3)~.
\end{equation}
Note that the above integral is same as in \eref{H_st_1}. Hence, from the same argument as before, one can conclude for a homogeneous medium that
\begin{equation}
\frac{dH_{2}(t)}{dt} = 0~.
\end{equation}
In the following section we present the case when the two point correlations are independent of velocity and time. 

\subsection{Velocity and time independent correlations}

Let us assume that $g_{2}$ is independent of momentum and time.   
Then, \eref{eq:A} reduces to 
\begin{eqnarray}
	\left[ \frac{\partial }{\partial t} + \vec{v}_{1} \cdot \vec{\nabla}_{r_1} + \vec{a}_{int}^{(1)}(1) \cdot \frac{\partial }{\partial \vec{v}_{1}} \right]~ \ln(f_{1}(1)) = 0~, \label{eq:1.38} 
\end{eqnarray}
since there is no velocity and time dependence in $g_{2}$. Hence \eref{eq:1.38} can be written as,
\begin{equation}
	\left[ \frac{\partial }{\partial t} + \vec{v}_{1} \cdot \vec{\nabla}_{r_1} \right]f_{1}(1) = -\vec{a}_{int}^{(1)}(1) \cdot \frac{\partial }{\partial \vec{v}_{1}} f_{1}(1)~,
\end{equation}
where,
\begin{eqnarray}
\vec{a}_{int}^{(1)}(1) &=& -\int \frac{\partial \phi_{12}}{\partial \vec{r}_{1}}~f_{1}(2)~g_{2}(1,2)~d^{3}(2) \nonumber \\[2mm]
&=& -\int \frac{\partial \phi_{12}}{\partial \vec{r}_{1}}~f_{1}(\vec{r}_{2},\vec{v}_{2},t)~g_{2}(\vec{r}_{1},\vec{r}_{2})~d^{3}(2) \nonumber \\[2mm]
\vec{a}_{int}^{(1)}(\vec{r}_{1},t) &=& -\int \rho(\vec{r}_{2},t)~ g_{2}(\vec{r}_{1},\vec{r}_{2})~ \frac{\partial \phi_{12}}{\partial \vec{r}_{1}}~d^{3}r_{2} \label{eq:1.40} \\[2mm]
	&=& \vec{a}_{int}^{(m)}(\vec{r}_{1},t)  \nonumber
\end{eqnarray}
In the above, mass density $\rho(\vec{r},t)$ is defined as,
\begin{equation}
\rho(\vec{r},t) = \int f_{1}(\vec{r},\vec{v},t) ~d^{3}v~,
\label{rho_def}
\end{equation}
and ~$\vec{a}_{int}^{(m)}(\vec{r}_{1},t)$~ refers to the mean-field force term in the Martys' kinetic equation, see~\cite{Mar_IJMPC1999,Mar_PhysA2006} for more details. Hence if we drop the momentum and time dependence in $g_{2}$, the first of the coupled equations, \eref{eq:A} reduces to the Martys kinetic equation,
\begin{eqnarray} 
	\frac{\partial{f_{1}}}{\partial{t}}&+&\vec{v}_{1}\cdot \frac{\partial{f_{1}}}{\partial{\vec{r}_{1}}}+\vec{a}_{ext}\cdot \frac{\partial{f_{1}}}{\partial{\vec{v}_{1}}}~=~\Omega_{1}~, \nonumber \\[2mm]
	\label{martyseq}
	\Omega_{1}~ &=& ~\int{\frac{\partial \phi (r_{12})}{\partial \vec{r}_{1}} \cdot \frac{\partial (f_{1}({1})~f_{1}({2})~g_{2})}{\partial \vec{v}_{1}}~d\vec{r}_{2}~d\vec{v}_{2}} \\[2mm]
	&=& ~\frac{\partial f_{1}({1})}{\partial \vec{v}_{1}}~\cdot~\int{\frac{\partial \phi (r_{12})}{\partial \vec{r}_{1}}~f_{1}({ 2})~g_{2}(\vec{r}_{1},\vec{r}_{2},t)~d\vec{r}_{2}~d\vec{v}_{2} } \nonumber \\[2mm]
	&=&~- \frac{\partial{f_{1}}}{\partial{\vec{v}_{1}}} \cdot \vec{a}_{int}(\vec{r}_{1},t)~. \nonumber
\end{eqnarray}
The second of the set of coupled equations, \eref{eq:B}, becomes
\begin{eqnarray}
	\left( \vec{v}_{1} \cdot \vec{\nabla}_{r_1}+ \vec{v}_{2} \cdot \vec{\nabla}_{r_2} \right)\ln(g_{2}&&(1,2))= \nonumber \\
	&&=\left( \vec{a}_{int}^{(1)}(1) - \vec{a}_{int}^{(2)}(1^{+}2) \right) \cdot \vec{\nabla}_{v_{1}} \ln(f_{1}(1)) + \nonumber \\
	&&+\left( \vec{a}_{int}^{(1)}(2) - \vec{a}_{int}^{(2)}(12^{+}) \right) \cdot \vec{\nabla}_{v_{2}} \ln(f_{1}(2))~.
	\label{eq:1.43a}
\end{eqnarray}
From \eref{eq:1.40} we know that $\vec{a}_{int}^{(1)}(1) \rightarrow \vec{a}_{int}^{(m)}(\vec{r}_{1},t)$, similarly  ~$\vec{a}_{int}^{(1)}(2) \rightarrow \vec{a}_{int}^{(m)}(\vec{r}_{2},t)$ and 
\begin{eqnarray}
	\vec{a}_{int}^{(2)}(1^{+}2) &=& -\int \frac{\partial \phi_{13}}{\partial \vec{r}_{1}} ~f_{1}(\vec{r}_{3},\vec{v}_{3},t) ~g_{2}(\vec{r}_{1},\vec{r}_{3}) ~g_{2}(\vec{r}_{2},\vec{r}_{3}) ~d^{3}r_{3}~ d^{3}v_{3}~, \nonumber \\[2mm]
	&=& -\int \rho(\vec{r}_{3},t) ~g_{2}(\vec{r}_{1},\vec{r}_{3}) ~g_{2}(\vec{r}_{2},\vec{r}_{3}) ~\frac{\partial \phi_{13}}{\partial \vec{r}_{1}} ~~d^{3}r_{3}~.
\end{eqnarray}	
From the above, we can write
\begin{eqnarray} 
\vec{a}_{int}^{(2)}(\vec{r}_{1}^{+},\vec{r}_{2},t) &&= \nonumber \\ &&=-\int \rho(\vec{r}_{3},t) ~g_{2}(\vec{r}_{1},\vec{r}_{3}) ~ h_{2}(\vec{r}_{2},\vec{r}_{3})~\frac{\partial \phi_{13}}{\partial \vec{r}_{1}}~d^{3}r_{3}~- \nonumber\\
&&~~~~~~~~-\int \rho(\vec{r}_{3},t) ~g_{2}(\vec{r}_{1},\vec{r}_{3}) ~\frac{\partial \phi_{13}}{\partial \vec{r}_{1}} ~d^{3}r_{3}~, 
\label{eq:1.44a}
\end{eqnarray}
where,
\begin{equation}
	h_{2}(\vec{r}_{2},\vec{r}_{3}) = \left\lbrace g_{2}(\vec{r}_{2},\vec{r}_{3}) - 1 \right\rbrace ~.
	\label{eq:1.45a}
\end{equation}
From \eref{eq:1.44a} and \eref{eq:1.45a}, we can write
\begin{eqnarray}
	\vec{a}_{int}^{(2)}(\vec{r}_{1}^{+},\vec{r}_{2},t) &=& -\int \rho(\vec{r}_{3},t) ~g_{2}(\vec{r}_{1},\vec{r}_{3}) ~h_{2}(\vec{r}_{2},\vec{r}_{3})  ~\frac{\partial \phi_{13}}{\partial \vec{r}_{1}} ~d^{3}r_{3}~+ \nonumber \\[2mm] &~& \hspace*{1cm} + \vec{a}_{int}^{(m)}(\vec{r}_{1},t)~,
	\label{eq:1.46a}
\end{eqnarray}
and a similar calculation will show that 
\begin{eqnarray}
	\vec{a}_{int}^{(2)}(\vec{r}_{1},\vec{r}_{2}^{+},t) &=& -\int \rho(\vec{r}_{3},t) ~h_{2}(\vec{r}_{1},\vec{r}_{3}) ~g_{2}(\vec{r}_{2},\vec{r}_{3})  ~\frac{\partial \phi_{13}}{\partial \vec{r}_{1}} ~d^{3}r_{3}~+ \nonumber \\[2mm] &~& \hspace*{1cm} + \vec{a}_{int}^{(m)}(\vec{r}_{2},t)~.
	\label{eq:1.47a}
\end{eqnarray}
In the steady state limit, take $f_{1}$ to be a Maxwellian, {\it i.e.,} $\displaystyle f_{1} \approx A e^{-\beta \frac{1}{2} m \vec{v}^{2}}$. Now, from \eref{eq:1.43a}, \eref{eq:1.46a} and \eref{eq:1.47a}, one can show that
\begin{equation}
	\vec{\nabla}_{r_{1}} \ln g_{2}(\vec{r}_{1},\vec{r}_{2}) = -\beta \phi_{12} -\beta \rho_{0} \int g_{2}(\vec{r}_{1},\vec{r}_{3})~h_{2}(r_{23})~\frac{\partial \phi_{13}}{\partial \vec{r}_{1}}~d^{3}r_{3}~,
	\label{eq:1.48a}
\end{equation}
where we assume the system is homogeneous, $\rho(\vec{r}_{3}) \approx \rho_{0}$. The above result is a truncated version of the BGY (Born-Green-Yvon) hierarchy, see~\cite{BGY_ref1,BGY_ref2}. Note that the BGY hierarchy relates $g_2$ to $g_3$ and so on. But the present scheme is equivalent to writing 
\begin{equation}
	g_{3}(1,2,3) \approx g_{2}(1,2)~g_{2}(2,3)~g_{2}(3,1)~.
\end{equation} 
From the above, which is called as ``{Kirkwood superposition approximation},''~\cite{KSA_ref}, the BGY hierarchy reduces to \eref{eq:1.48a}.

\section{Conclusions}

In this paper we propose a new closure scheme for BBGKY hierarchy by approximating $f_{3}$ in terms of its lower order distribution functions and two point correlation functions. This resulted in a set of coupled kinetic equations for $f_{1}(1,t)$ and $g_{2}(1,2,t)$. Since the momentum and time dependence of $g_{2}$ is retained, it should provide a more reliable estimate of time evolution of $f_{1}$ compared to earlier models~\cite{Mar_IJMPC1999,Mar_PhysA2006,Siva_JSP2014}. We have also shown that the present model reduces to ~\cite{Mar_IJMPC1999,Siva_JSP2014} for velocity independent correlations. Here, we have shown analytically that the current closure scheme respects time symmetry, \textit{i.e.}, $\displaystyle \frac{dH(t)}{dt}=0$, for homogeneous density. 

\section*{Acknowledgments}

IR would like to acknowledge financial support provided by the Centre for Modeling, Simulation and Design (CMSD), University of Hyderabad.


\section*{References}
\begin{thebibliography}{99}
%
\bibitem{BBGKY_ref1}{Klimontovich Y L 1995 \textit{Statistical Theory of Open Systems, vol 1: A Unified Approach to Kinetic Description of Processes in Active Systems} (Kluwer Academic Publishers-Plenum Publishers)} 
\bibitem{BBGKY_ref2}{Bogolyubov N N 1962 \textit{Problems in Dynamical Theory in Statistical Physics (in Russian); translation reprinted in: Studies in Statistical Mechanics, vol 1}, Edited by J. deBoer and G. E. Uhlenbeck (New York, Wiley)}
\bibitem{XuSucBogh_MCS2006}{Xu A, Succi S and Boghosian B M 2006 \textit{Mathematics and Computers in Simulation} \textbf{72} 249-252}
\bibitem{Mar_IJMPC1999}{Martys N S 1999 \textit{International Journal of Modern Physics C} \textbf{10} 1367-1382}
\bibitem{Mar_PhysA2006}{Martys N S 2006 \textit{Physica A: Statistical Mechanics and its Applications} \textbf{362} 57 - 61}
\bibitem{Siva_JSP2014}{Chari S S N, Inguva R and Murthy K P N 2014 \textit{Journal of Statistical Physics} \textbf{157} 113-123}
\bibitem{BGY_ref1}{Born M and Green H S 1949 \textit{A General Kinetic Theory of Liquids} (Cambridge University Press, England)}
\bibitem{BGY_ref2}{Green M S 1956 \textit{J. Chem. Phys.} \textbf{25} 836}
\bibitem{KSA_ref}{Kirkwood J G 1947 \textit{J. Chem. Phys.} \textbf{15} 72}
\bibitem{Jaynes_1}{Jaynes E T 1957 \textit{Phys. Rev.} \textbf{106} 620-630}
\bibitem{Jaynes_2}{Jaynes E T 1957 \textit{Phys. Rev.} \textbf{108} 171-190}
\bibitem{Mitchell_thesis}{Mitchell W C 1957 Statistical mechanics of thermally driven systems \textit{PhD Thesis} Washington University}

\end{thebibliography}
\end{document}